\documentclass[angeo]{copernicus}

\usepackage{color}
\usepackage{url}

\newcommand{\zeroindexLarge}{0}
\newcommand{\zeroindexSmall}{{\rm o}}
\newcommand{\Vzero}{V_\zeroindexLarge}
\newcommand{\nzero}{n_\zeroindexSmall}
\newcommand{\epsilonzero}{\epsilon_\zeroindexSmall}

\newcommand{\MARKI}[1]{#1}
\newcommand{\MARKIII}[1]{#1}

\begin{document}


\title{Boltzmann electron PIC simulation of the E-sail effect}

\author{P.~Janhunen}

\affil{Finnish Meteorological Institute, POB-503, FI-00101, Helsinki, Finland}


\runningtitle{E-sail Boltzmann PIC simulation}

\runningauthor{P.~Janhunen}

\correspondence{Pekka Janhunen\\ (pekka.janhunen@fmi.fi)}

\received{}
\pubdiscuss{} 
\revised{}
\accepted{}
\published{}


\firstpage{1}

\maketitle  

\begin{abstract}
The solar wind electric sail (E-sail) is a planned in-space propulsion
device that uses the natural solar wind momentum flux for spacecraft
propulsion with the help of long, charged, centrifugally stretched
tethers. The problem of accurately predicting the E-sail thrust is
still somewhat open, however, due to a possible electron population
trapped by the tether. Here we develop a new type of particle-in-cell
(PIC) simulation for predicting E-sail thrust. In the new simulation,
electrons are modelled as a fluid, hence resembling \MARKI{hybrid} simulation,
but in contrast to normal hybrid simulation, the Poisson equation is
used as in normal PIC to calculate the self-consistent electrostatic
field. For electron-repulsive parts of the potential, the Boltzmann
relation is used. For electron-attractive parts of the potential we
employ a power law which contains a parameter that can be used to
control the number of trapped electrons. We perform a set of runs
varying the parameter and select the one with the smallest number of
trapped electrons which still behaves in a physically meaningful way
in the sense of producing not more than one solar wind ion deflection
shock upstream of the tether. By this prescription we obtain thrust
per tether length values that are in line with earlier estimates,
although somewhat smaller. We conclude that the Boltzmann PIC
simulation is a new tool for simulating the E-sail thrust. This tool
enables us to calculate solutions rapidly and allows to easily study
different scenarios for trapped electrons.
\end{abstract}


\introduction  

The electric solar wind sail (electric sail, E-sail) is a planned
device for producing interplanetary spacecraft propulsion from the
natural solar wind by a set of centrifugally stretched thin metallic
tethers that are kept artificially at a high positive potential
\citep{RSIpaper}. In order to engineer E-sail devices in detail, one
should predict the magnitude of the Coulomb drag that the flowing
solar wind exerts on the charged tether. Thus far this problem has
been studied using particle-in-cell (PIC) simulations
\citep{paper2,paper6,ASTRONUM2011}\MARKI{, Vlasov simulations \citep{SanchezArriagaAndPastorMoreno2014}} and other methods
\citep{SanmartinEtAl2008,SanchezTorres2014}. However, a fully
satisfactory way of estimating E-sail thrust has not yet emerged,
except for negative polarity tethers \citep{paper24}, which are
however more suitable to use in low Earth orbit (LEO) as a deorbiting
plasma brake device \citep{Plasmabrake} than in the solar wind
\citep{paper3}. Independent laboratory measurements of the width of the electron
sheath around a positively biased tether placed in streaming plasma
were made by \citet{SiguierEtAl2013}. The laboratory results were
produced in a plasma mimicking conditions in LEO and they are in good
agreement with PIC thrust predictions \citep{Koln2014}.

In this paper we develop a new variant of the PIC simulation of
\citet{ASTRONUM2011} which treats electrons as a fluid obeying the
Boltzmann relation, with two additional tricks to be detailed
below. The motivation is to have a code which is relatively fast and
easy to run and which makes it possible for the user to control the
assumed amount of trapped electrons in a simple way. We present the
simulation code and use it to derive a thrust estimate in a case which
has relevance to the E-sail. Making more comprehensive set of runs
with different voltages and in different solar wind conditions is
outside of the scope of the paper.

\section{Physics of tether Coulomb drag}

The negative polarity Coulomb drag effect can be successfully
simulated by a PIC code \citep{paper24}. The negative polarity effect
is less challenging to simulate than the positive polarity one,
because in the negative polarity case, trapped particles do not
form. Electrons are not trapped because they are repelled by the
tether's negative potential structure. Ions (protons) are also not
trapped, because in relevant cases the surrounding plasma flow is
supersonic so that in a coordinate frame where the tether is
stationary, ions enter the potential structure with significant
kinetic energy and hence cannot be easily trapped by the potential.

In the positive polarity case which is the subject of the present
paper, electrons may become trapped because the tether attracts them
and because the electron flow is subsonic so that there are some
electrons which enter the potential structure with nearly zero initial
energy.  The issue of trapped electrons was identified by
\citet{paper2} and simulations were later made where chaotisation of
trapped electron orbits when scattering at the end of the tether and
at the spacecraft was explicitly included in the PIC code, as well as
occasional removal of trapped electrons due to collisions with the
tether \citep{ASTRONUM2011}. However, in that calculation the
timescale where trapped electrons were removed was made much faster
than in reality, in order to complete the simulation run in a
reasonable time. Basically, processes controlling the size of the
trapped electron population are poorly known, because we think that
they happen at longer timescales than what is accessible by PIC
simulation.

Consider a 2-D positive potential structure which thus forms a
negative potential energy well (attractive structure) for
electrons. If the potential is stationary and hosts no trapped
electrons and if the ambient electron distribution is a non-drifting
Maxwellian and the plasma is tenuous enough to be collisionless, then
it can be shown using Liouville's theorem that the local electron
density can nowhere exceed the background plasma density
\citep{LaframboiseAndParker1973,paper6}. In fact, under the stated
conditions the electron density is typically \MARKIII{significantly} less than the
background density in most regions covered by the attractive potential
structure. Physically the result is due to the fact that although the
tether attracts electrons and therefore focusses them to move through its vicinity,
the electron density remains low because accelerating the electron
reduces the time that the electron spends inside the attractive potential
structure. Conservation of the electron's originally nonzero angular momentum also limits the minimum
distance that the electron can have from the tether.

That the electron density nowhere exceeds the background density leads
to an interesting dilemma (\'Eric Choini\`ere, private communication)
that is associated with trapped electrons because the ion density, on
the other hand, must be higher than the background density inside at
least in some regions of the upstream side ion deflection shock. Then,
a significant positive charge density can build up at the upstream ion
deflection region, which spreads out the electron-attractive potential
structure and thus decreases the local electron density further which,
unless some process intervenes, makes the charge imbalance worse. What
exactly happens remains not well known, but conceivably, instabilities
which are able to scatter electrons might develop and take place until
enough trapped electrons have been formed so that proper charge
neutralisation in the ion deflection shock region has been restored. A
simple approach to model this kind of general behaviour may be to
reduce the number of trapped electrons in the simulation until the
result becomes nonphysical in some way or another and postulate that
the most realistic run is the physical run with least amount of
trapped electrons. This is what the simulation presented next attempts
to accomplish.

\section{Simulation code}

We use a special version of electrostatic PIC where electrons are not
modelled as particles, but as a fluid whose local density $n_e(x,y)$
is an assumed function of the local potential $V(x,y)$. When the
potential repels electrons (i.e.~when $V(x,y)<0$), the Boltzmann
relation is assumed,
\begin{equation}
n_e(x,y) = \nzero \exp\left[\frac{eV(x,y)}{T_e}\right],\quad \mathrm{when}\ V(x,y)<0
\label{eq:repulsive}
\end{equation}
where $\nzero$ is the background plasma density and $T_e=10$ eV is the
background electron temperature in energy units. The Boltzmann
relation is supposedly a good approximation for the density of
particles that try to penetrate thermally into a repulsive potential
structure.

For attractive parts of the potential, the electron density depends on
the number and distribution function of trapped electrons and thus
extra assumptions are needed. That one needs such assumptions is
natural because we are not modelling electron dynamics using first
principles. We want to have a functional relationship which contains a
parameter that can be used to tune the number of trapped electrons in
a simple way. 

In this paper we use the following functional form for electron
density in attractive parts of the potential:
\begin{equation}
n_e(x,y) = \nzero \left[1+\frac{e V(x,y)}{T_e}\right]^\nu
\quad\mathrm{when}\ V(x,y)\ge 0
\label{eq:attractive}
\end{equation}
where $\nu>0$ is the trapped electron control parameter.

\begin{figure}[t]
\includegraphics[width=8.4cm]{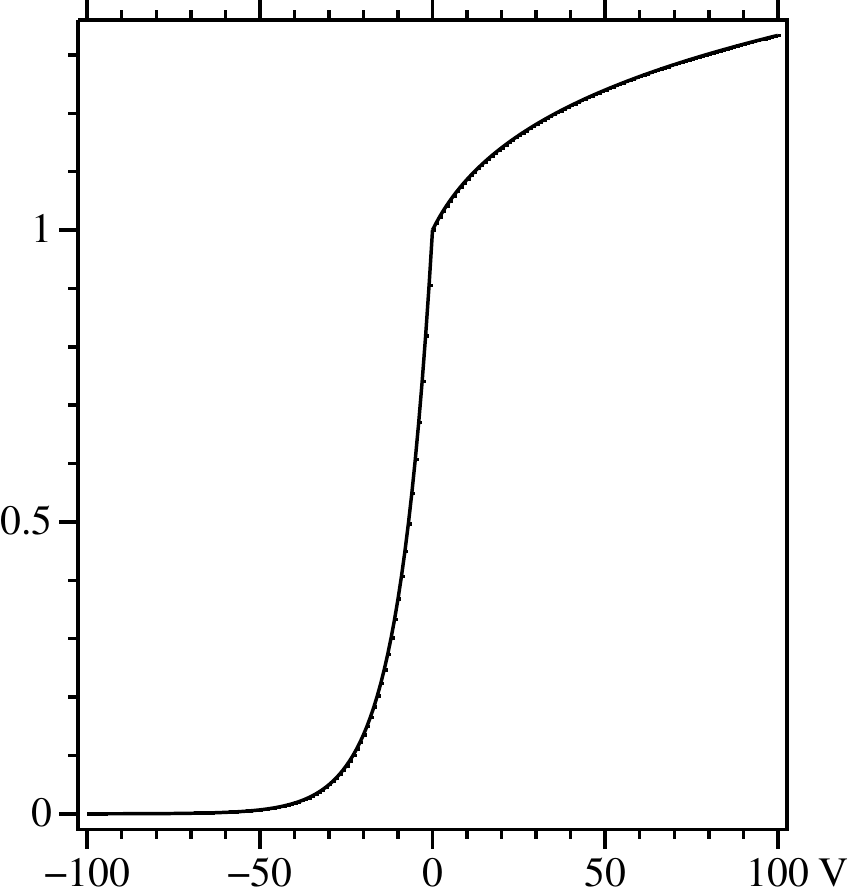}
\caption{Local relative electron density $n_e(x,y)/\nzero$ as function of
  local potential $V(x,y)$ for $\nu=0.12$, by Eqs.~(\ref{eq:repulsive})
  and (\ref{eq:attractive}). The electron temperature is $T_e=10$ eV.}
\label{fig:necurve}
\end{figure}

In Figure \ref{fig:necurve} we show how the local electron density
depends on the local potential according to Eqs.~(\ref{eq:repulsive})
and (\ref{eq:attractive}).

The code that we use in this paper is a descendant of a PIC code that
we used earlier \citep{paper2,ASTRONUM2011,paper24}. It is an
electrostatic code with linear particle weighting and additional
grid-level smoothing which is implemented in the Fourier domain. The
code supports 2-D and 3-D simulations and optionally includes a
constant external magnetic field. The Poisson solver is implemented by
fast Fourier transform and the code is parallelised with OpenMP
threads and the Message Passing Interface (MPI) library. The code is
also vectorised, so that for a processor with 256-bit wide AVX2
instruction set, each processor core calculates eight particle updates
simultaneously. For particle coordinates and velocities, 32-bit
floating point accuracy is used. To avoid loss of accuracy, particle
coordinates are expressed relative to the centroid of the grid cell
where the particle resides. Particles are stored in lists belonging to
grid cells which enables fast accumulation of the charge density (no
irregular memory addressing patterns needed) and enables good data
locality and cache friendliness. When a particle crosses a grid cell
boundary, it is removed from the old list and entered into a new one.

An additional trick is needed to avoid solutions that look unphysical
and oscillate drastically in time.  Namely, the electron density must not
react instantaneously to changed electric potential, but with some
delay. If $n_e^*(x,y)$ is the preliminary instantaneous electron
density computed by Eqs.~(\ref{eq:repulsive}) and
(\ref{eq:attractive}), then we propagate the real electron density
$n_e(x,y)$ by the differential equation
\begin{equation}
\frac{\partial n_e}{\partial t} = \frac{n_e^*-n_e}{\tau}
\label{eq:diffeq}
\end{equation}
where $\tau$ is a time constant. The solution $n_e(t)$ of
(\ref{eq:diffeq}) is effectively a low-pass filtered version of
$n_e^*(t)$ where frequencies above $\sim 1/\tau$ are filtered out. We
have tested different values of $\tau$ and found that in order to
produce good results (in the sense of avoiding unphysical
oscillations), $\tau$ should be proportional to the ion plasma
oscillation period, because shorter timescale phenomena are not
included in a model with electron fluid. Specifically, we use
\begin{equation}
\tau = C\,\frac{1}{\omega_{pi}},\quad C=1.5
\label{eq:tau}
\end{equation}
where $\omega_{pi}$ is the ion plasma frequency
\begin{equation}
\omega_{pi} = \sqrt{\frac{\nzero e^2}{\epsilonzero m_i}}.
\label{eq:omegapi}
\end{equation}
By numerical experimentation, $C\le 1$ produces unphysical
oscillations which disappear if $C>1$. We use $C=1.5$ to have some
extra margin of safety. On the other hand there is no reason to
increase $C$ further.

In summary, the code pushes ions forward by their equations of motion
(in this paper with $\mathbf{B}$=0),
\begin{eqnarray}
\label{eq:eqs_of_motion}
\frac{d \mathbf{v}_i(t)}{dt} &=& \MARKIII{\frac{e}{m_i}} \left[\mathbf{E}(\mathbf{x}_i(t))
+ \mathbf{v}_i(t)\times\mathbf{B}\right]\nonumber \\
\frac{d \mathbf{x}_i(t)}{dt} &=& \mathbf{v}_i(t),
\end{eqnarray}
accumulates ion density $n_i$ from the ion positions $\{\mathbf{x}_i(t)\}$
\begin{equation}
n_i({\bf x},t) = \sum_i S(\mathbf{x}-\mathbf{x}_i(t))
\end{equation}
where $S$ is the linear area weighing function
\citep{BirdsallAndLangdon}. Then the code applies Fourier domain filtering
to smooth the ion density spatially, calculates the provisional electron density
$n_e^*$ by Eqs.~(\ref{eq:repulsive}) and (\ref{eq:attractive}),
updates the electron density $n_e$ by Eq.~(\ref{eq:diffeq}), computes
the charge density $\rho_e = e(n_i-n_e)$, solves the Poisson equation
\begin{equation}
-\nabla^2\phi(\mathbf{x}) = \frac{1}{\epsilonzero}\,\rho_e(\mathbf{x}),
\end{equation}
computes the gridded electric field $\mathbf{E}(\mathbf{x})$ from $\phi(\mathbf{x})$ by finite
differencing and repeats the process for the next timestep.

The particle equations of motion (\ref{eq:eqs_of_motion}) are
discretised in time using the second order accurate leapfrog method as
is common in momentum conservative explicit PIC simulations \citep{BirdsallAndLangdon}. Equation (\ref{eq:diffeq}) is
discretised by forward Euler method. This is equivalent to linearly
mixing the portion $\Delta t/\tau$ of $n_e^*$ into $n_e$ at each
timestep, where $\Delta t$ is the length of the timestep. Table
\ref{tab:simparams} summarises the simulation parameters.

\MARKIII{
The timestep $\Delta t$ in our code is $\sim 50$ times longer than
what it would be in a full electron-ion PIC code using the same grid
resolution. In addition to that, it seems that the code can produce
useful results with a smaller number of ions per grid cell than a full
PIC code so that the improvement factor in practical computational
efficiency is probably several hundred. The latter property
is probably due to averaging over timescale $\tau\approx 210\,\Delta t$
which effectively reduces the numerical noise.
}

\section{Results}

First we made a family of runs at low resolution, varying the trapped
electron parameter $\nu$ between 0.16 and 0.08. The grid spacing was
10 m and the box side length 1.92 km so that the grid was $192\times
192$. The timestep was 4 $\mu$s, corresponding to 2500 km/s maximum
signal speed, resulting in Courant-Friedrichs-Lewy (CFL) parameter of
0.16 with respect to the used solar wind speed of 400 km/s (Table
\ref{tab:simparams}). The solar wind plasma density is 7.3 cm$^{-3}$
and for simplicity, no interplanetary magnetic field is used. The ion
and electron temperature is 10 eV. Pure proton plasma is used as the
solar wind. The plasma parameters correspond to average solar wind
properties at 1 au distance from the sun.

The real solar wind also has typically 10 nT magnetic field. However,
the resulting background electron Larmor radius of $\sim$1 km is much
larger than the diameter of the electron sheath $\sim$100-200 m. Thus
the magnetic field is expected to play only a minor role as far as the
magnitude of the E-sail thrust is concerned, so that assuming zero
magnetic field is not a bad first approximation.

For each run we determined the thrust by two complementary methods,
``Mom'' and ``Coul''. In method Mom we keep track of the momentum $x$
component of all particles entering and leaving the simulation box and
deduce that the momentum which is left in the box is the one which
gets transferred to the tether and the spacecraft. In method Coul we
evaluate the negative gradient of the plasma potential at the tether's
location and use Coulomb's law, $dF/dz=\lambda E$ where $dF/dz$ is the
thrust per tether length, $\lambda$ is the tether's linear charge
density and $E$ is the electric field caused by the plasma and
evaluated at the tether. The line charge $\lambda$ is given as
an input parameter \MARKI{(in the runs performed, $\lambda=8.64\cdot10^{-8}$
As/m),} and the corresponding tether voltage $\Vzero$ is
calculated from the run results after the end of the run by summing
the tether's vacuum potential and the potential created by the
plasma. Because the plasma potential varies somewhat from run to run,
the resulting tether voltage $\Vzero$ is also slightly different in
each run. For each run, we made a qualitative visual check of the
result to determine if the solution was physically reasonable in the
sense that it contained not more than one ion deflection shock.

We then made an identical family of runs with halved grid spacing of 5
m. The computing time is about 8 times longer than in the
low resolution set. Results of both families of runs are summarised in
Table \ref{tab:runs}.

\begin{figure*}[t]
\includegraphics[width=0.49\textwidth]{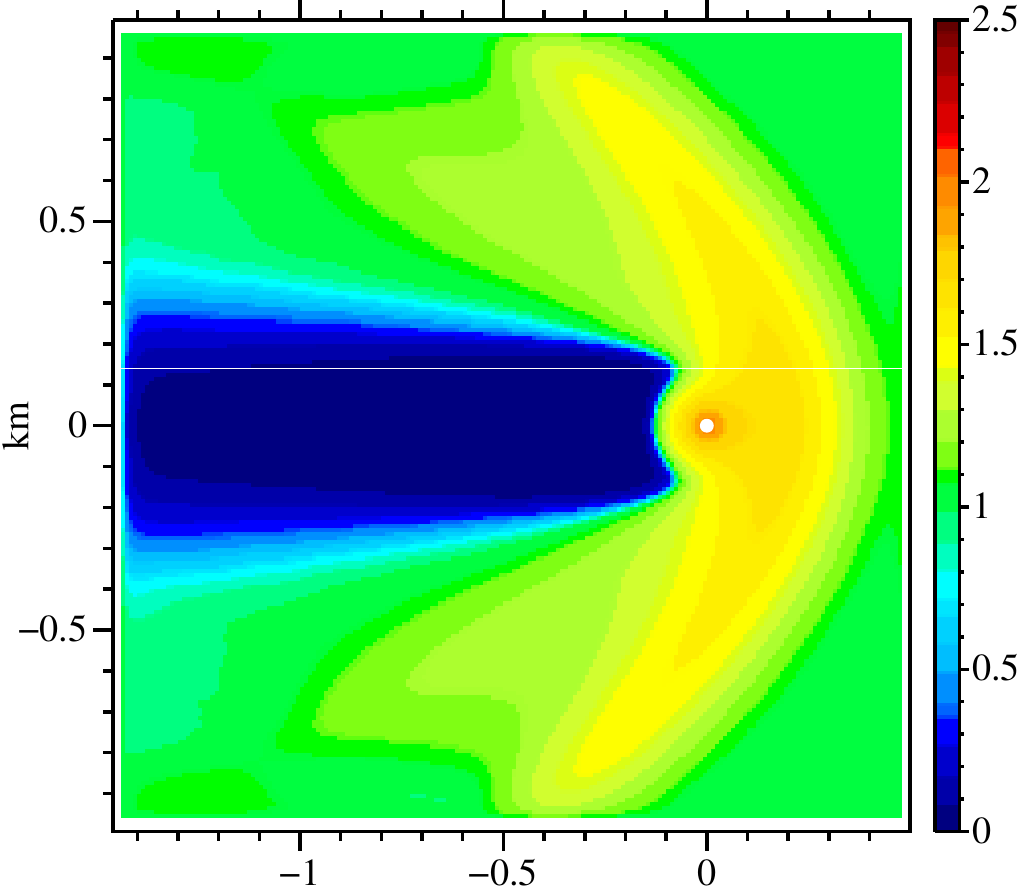}
\includegraphics[width=0.49\textwidth]{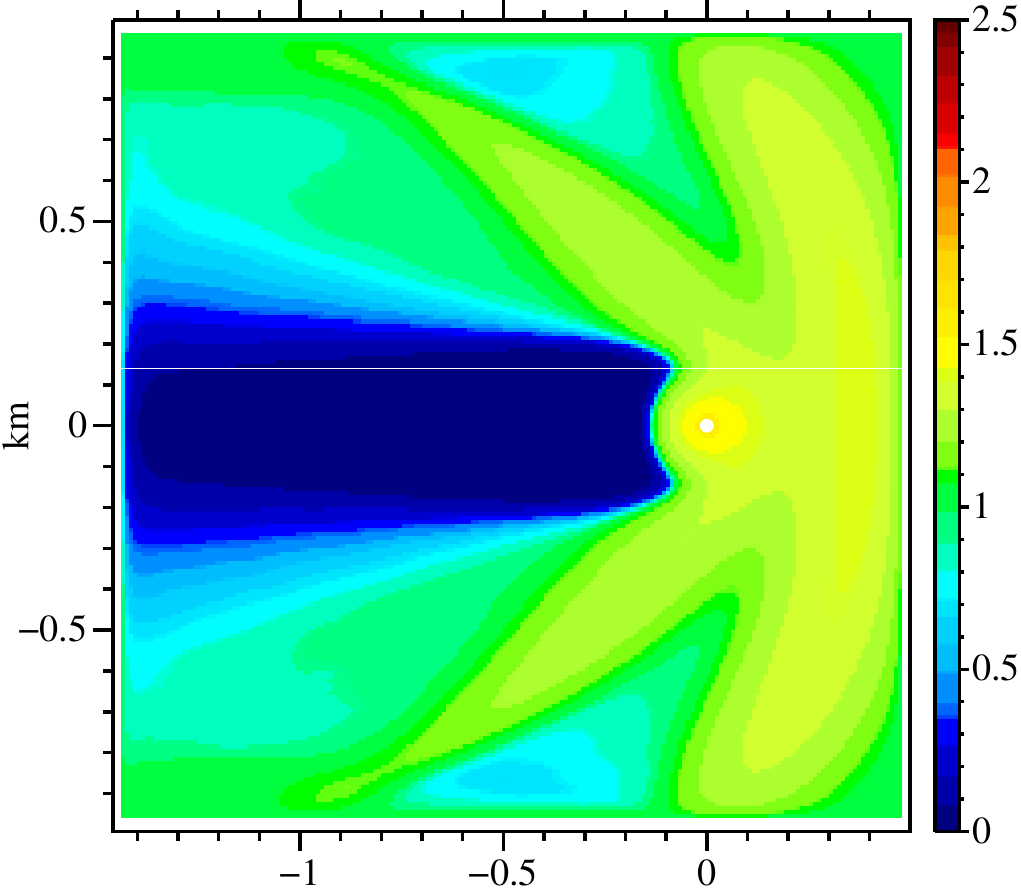}
\includegraphics[width=0.49\textwidth]{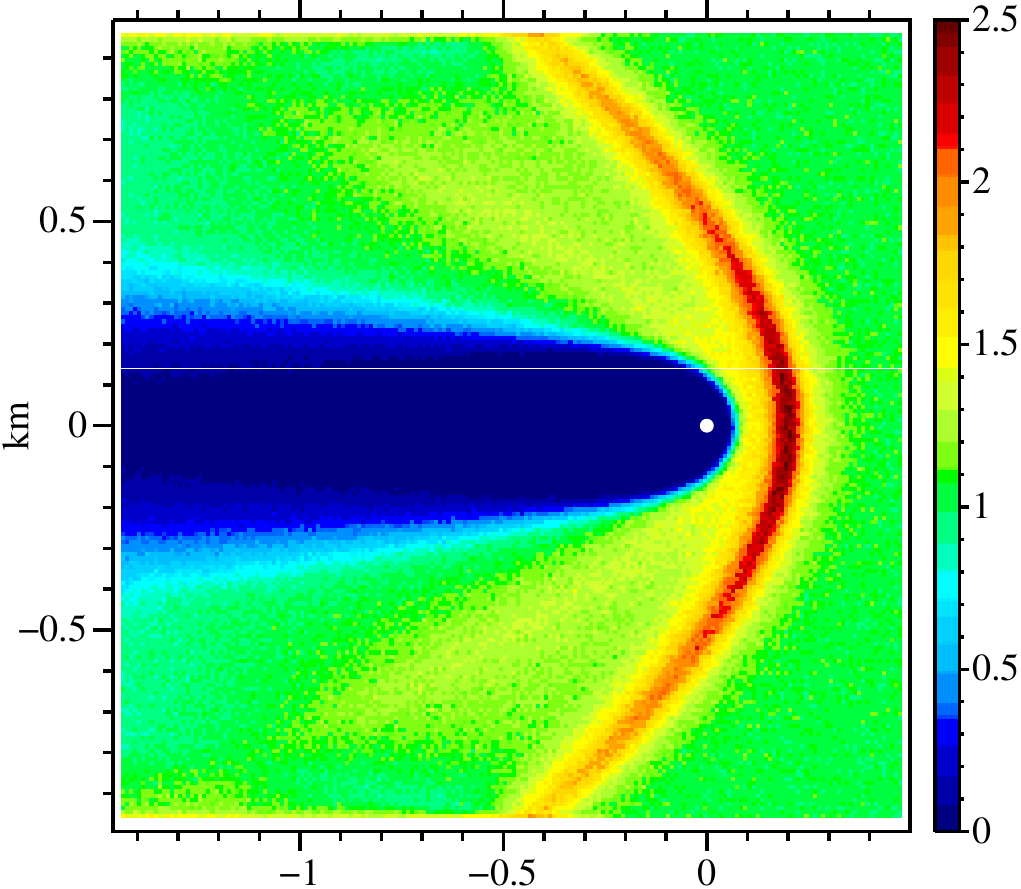}
\includegraphics[width=0.49\textwidth]{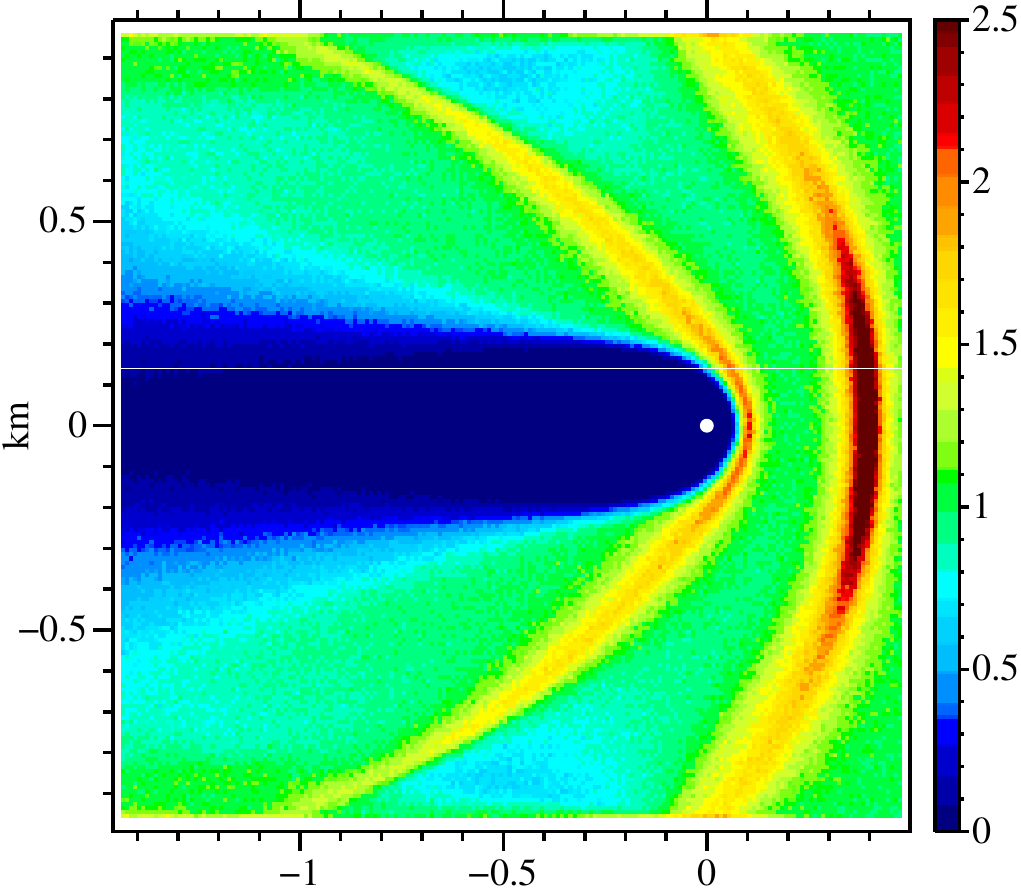}
\includegraphics[width=0.49\textwidth]{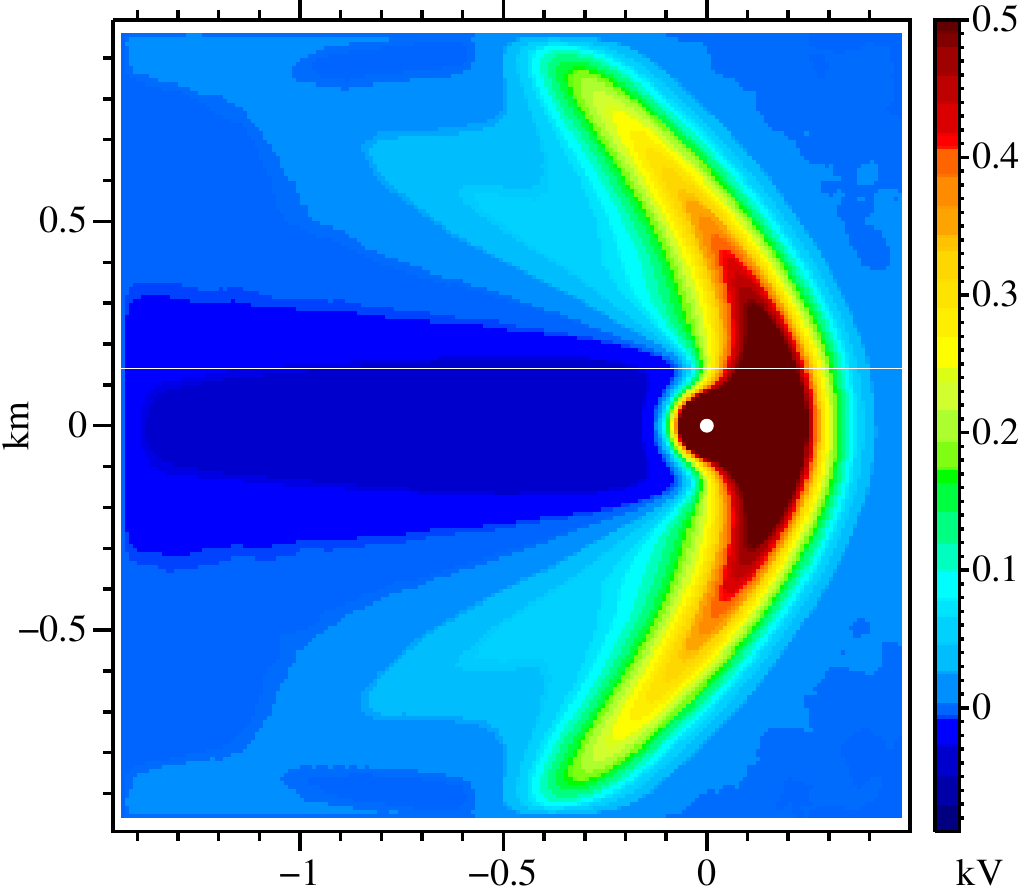}
\includegraphics[width=0.49\textwidth]{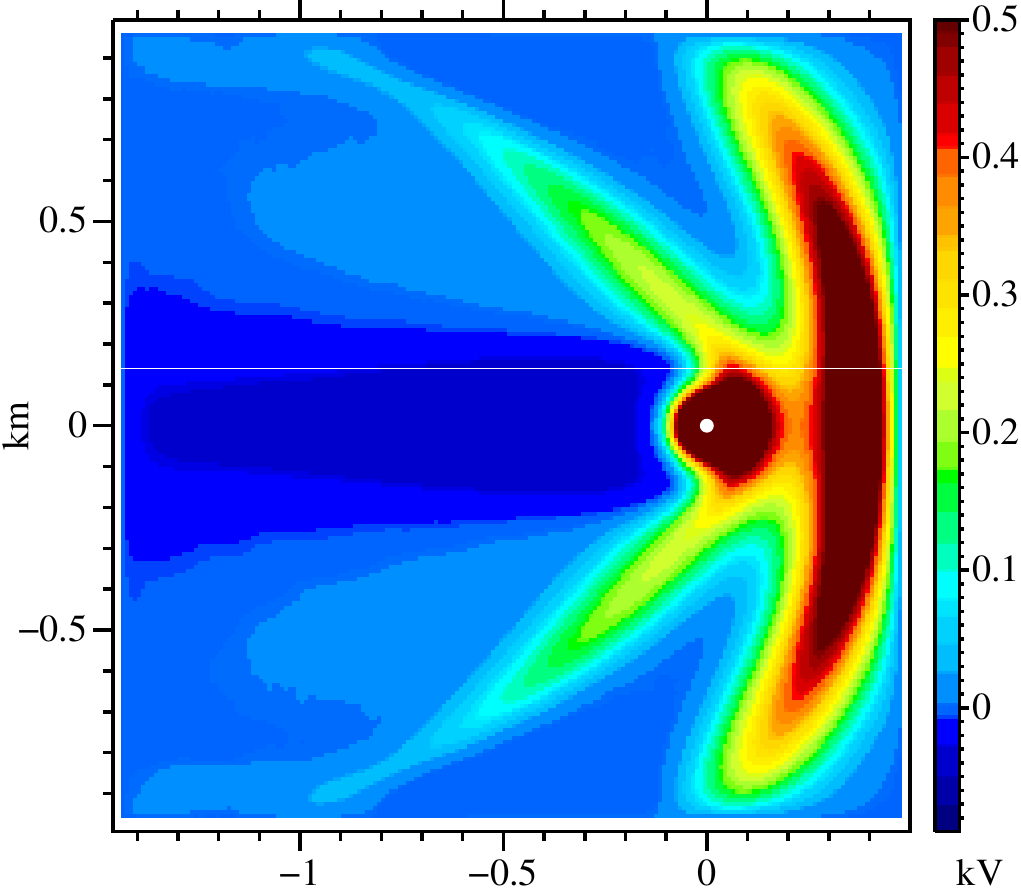}
\caption{Physically realistic run (left, $\nu=0.12$)
  and non-physical run (right, $\nu=0.08$) which exhibits double ion
  deflection shock. The panels are electron density (top), ion density (middle)
  and electric potential (bottom). Densities are relative to
  $\nzero$. White dot shows tether position at the origin.}
\label{fig:runs}
\end{figure*}

Run $\nu=0.12$ is selected as the baseline because our visual
inspection indicates that $\nu=0.12$ is the smallest value of $\nu$
for which we get not more than one ion deflection shock.  Figure
\ref{fig:runs} shows the electron and ion density and the total
potential in the baseline $\nu=0.12$ run as well as in another run
where $\nu=0.08$ and which exhibits two ion shocks.  In addition, in
cases where two shocks emerge in the solution, the shocks also tend to
slowly propagate upstream so that full convergence is not reached. The
high resolution run data are used, but for viewing speed convenience
the spatial grid resolution has been halved by neighbour averaging
before plotting in Fig.~\ref{fig:runs}. The solar wind arrives from
the right in Fig.~\ref{fig:runs}.

In Fig.~\ref{fig:runs} one clearly sees that the run $\nu=0.12$ looks
physically meaningful and well-behaved, while run $\nu=0.08$ exhibits
two shocks and is thus physically not acceptable. Similar visual
judgement can be made also for the other performed runs, and
information presented for each run in Table \ref{tab:runs} was
compiled in this way.

\begin{figure}[t]
\includegraphics[width=8.4cm]{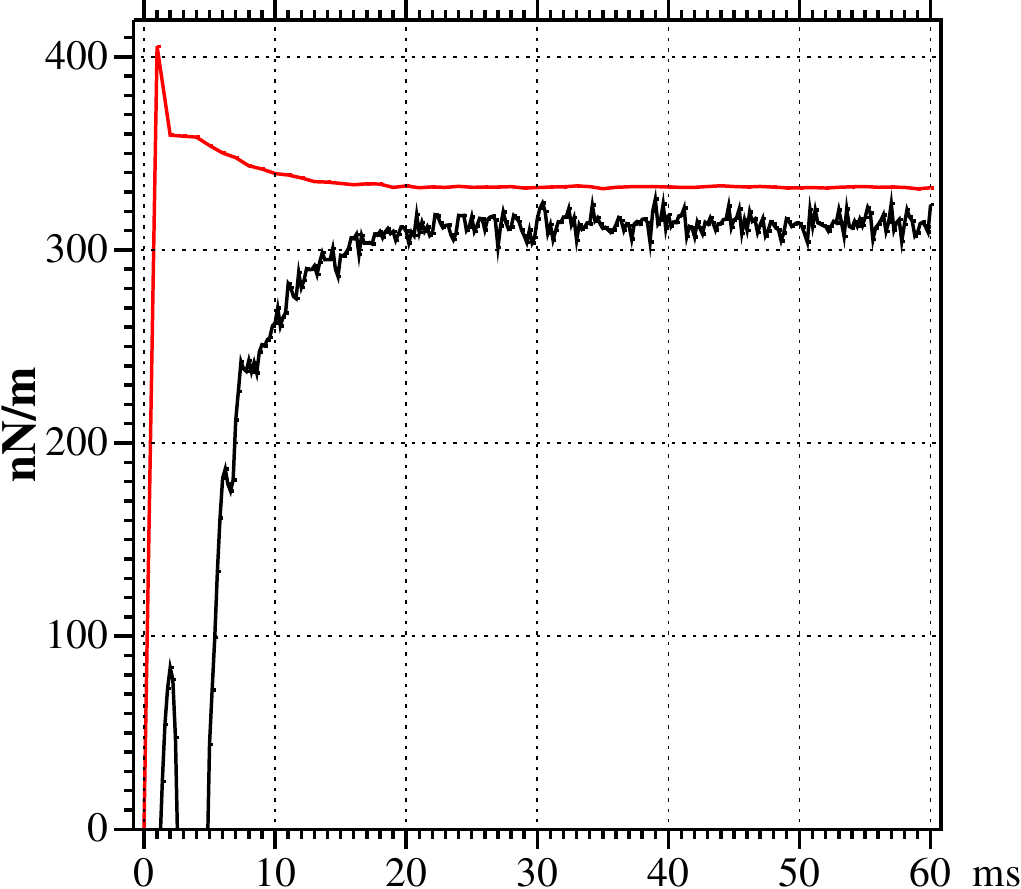}
\caption{Thrust history in $\nu=0.12$ run: method Mom (black) and method
  Coul (red).}
\label{fig:hist}
\end{figure}

Figure \ref{fig:hist} shows the thrust history of the $\nu=0.12$ run
(high resolution version). After natural initial transients have died
out, the Coul and Mom methods are in good agreement with each other
regarding the thrust per length produced by the E-sail. \MARKI{By
construction, the Mom method reacts to changes slower than Coul, and
during the first $\sim 10$ ms both methods give unreliable results.}

\conclusions[Discussion and conclusions]  

Using Boltzmann electrons in PIC simulation with two additional tricks
enables us to rapidly compute reasonable looking solutions for the
positive polarity Coulomb drag problem, applicable to the E-sail in
the solar wind. The first trick is that while the Boltzmann relation
(\ref{eq:repulsive}) is a natural choice for the electron density in
repulsive parts of the potential, in the attractive parts we use an a
power law relationship (\ref{eq:attractive}) containing one free
parameter that can be used to control the amount of trapped
electrons. The second trick is that to avoid spurious temporal
oscillations, the electron density is not updated instantaneously, but
frequencies higher than $\sim \omega_{pi}$ are filtered out by
Eq.~(\ref{eq:diffeq}).

The calculated solutions depend on parameter $\nu$ that must be
specified by the user and that parametrises the number of trapped
electrons. Once $\nu$ is fixed, the simulation produces a prediction
for the thrust per tether length. We calculated the thrust per length
by two complementary methods and found good mutual agreement.  For too
small values of $\nu$ the result looks unphysical, however, because it
contains two ion deflection shocks. Our strategy is that by finding
the smallest value of $\nu$ which gives a physically reasonable single
shock solution, we obtain a solution which is expectedly the best
approximation to reality within this framework.

For voltage $\Vzero$ of 17.7 kV and nominal solar wind at 1 au
(density 7.3 cm$^{-3}$, speed 400 km/s), the thrust predicted by the
framework is $\sim 320$ nN/m. In our earlier paper \MARKIII{\citep[Figure 12]{paper6}} we arrived at
estimate $500$ nN/m at $\Vzero=20$ kV \MARKIII{if
  there are no trapped electrons, 400 nN/m if trapped density equals
  background density, and 320 nN/m if the average trapped density in
  the sheath is four times
  higher than background}. If one
assumes that the thrust is approximately linearly proportional to
$\Vzero$, \MARKIII{after scaling} to 20 kV the new estimate becomes $\sim 360$ nN/m.
\MARKIII{The result is thus in quantitative agreement with \citet{paper6}
  when one takes into account that the average trapped density
  is $\sim 1.5$ in the new simulation. In summary,} if one
assumes that the number of trapped electrons increases until the
solution has only one ion deflection shock, then one ends up with
$\sim 360$ nN/m thrust estimate for 20 kV and average 1 au solar wind.

One must be somewhat cautious when interpreting the numerical result,
because the electron Boltzmann simulation in any case contains less
physics than the earlier full PIC simulations, and, for example, the
particular functional form (\ref{eq:attractive}) used to describe the
electron density in attractive parts of the potential was chosen
mostly by mathematical convenience. In any case, the Boltzmann
simulation is a new addition in the toolbox of the E-sail thrust
analyst. Its benefits are fast computation compared to traditional PIC
and easy possibility to test different assumptions concerning trapped
electrons.




\begin{acknowledgements}
The author thanks Petri Toivanen for reading the manuscript and making
a number of useful suggestions, \MARKIII{and \'Eric Choini\`ere for teaching the
trapped electron dilemma to the author}.
\end{acknowledgements}




\clearpage

\clearpage


\begin{table}[t]
\caption{Simulation parameters (high resolution runs). For
  definition of tether electric radius $r_w^*$, see \citet{paper2}.}
\begin{tabular}{lcr}
\tophline
Parameter & Symbol& Value \\
\middlehline
Grid size    & & $384 \times 384$ \\
Grid spacing & $\Delta x$ & $5$ m \\
$X$ domain & & -1.44 .. 0.48 km \\
$Y$ domain & & -0.96 .. 0.96 km \\
Timestep     & $\Delta t$ & $2 \mu$s \\
Run duration & $t_{\rm max}$ & $60$ ms \\
Number of timesteps & & 30000 \\
Ions per cell & $N_0$ & 100 (in plasma stream) \\
Number of particles & & $\sim 15\cdot 10^6$ \\
Plasma density & $\nzero$ & $7.3$ cm$^{-3}$ \\
Ion mass     & $m_i$ & 1 amu (protons) \\
Plasma drift & $v$ & 400 km/s \\
Electron temperature & $T_e$ & 10 eV \\
Ion temperature& $T_i$ & 10 eV \\
Electron Debye length & $\lambda_\mathrm{De}$ & 8.7 m \\
Tether voltage & $\Vzero$ & $\sim$17.7 kV \\
Tether electric radius & $r_w^{*}$ & 1 mm \\
\MARKI{Line charge density} & \MARKI{$\lambda$} & \MARKI{$8.64\cdot 10^{-8}$ As/m} \\
\bottomhline
\end{tabular}
\label{tab:simparams}
\end{table}


\begin{table*}[t]
\caption{Performed runs. Thrust per tether length in nN/m by Mom and
  Coul methods, NS=number of shocks. The physically preferred run $\nu=0.12$ is
  singled out by horizontal lines.}
\begin{tabular}{lll lll lll}
\tophline
       & \multicolumn{3}{c}{Low resolution runs} & \multicolumn{3}{c}{High resolution runs} & \multicolumn{2}{l}{} \\
\middlehline
$\nu$  & Mom   & Coul   & $\Vzero$/kV  & Mom   & Coul   & $\Vzero$/kV  & NS & Remarks\\
\middlehline
0.16   & 301.1 & 319.0 & 17.601 & 303.6 & 319.1  & 17.578 & 1 &\\
0.15   & 311.5 & 336.6 & 17.615 & 315.6 & 331.3  & 17.627 & 1 &\\
0.14   & 322.3 & 348.6 & 17.667 & 327.3 & 343.7  & 17.675 & 1 &\\
0.13   & 328.1 & 352.3 & 17.733 & 330.1 & 347.9  & 17.709 & 1 &Thrust local max\\
\middlehline
0.12   & 312.7 & 340.6 & 17.745 & 313.7 & 332.8  & 17.725 & 1 & Smallest $\nu$ with NS=1\\
\middlehline
0.11   & 296.0 & 328.5 & 17.751 & 300.7 & 319.8  & 17.749 & 1+ & \\
0.10   & 298.7 & 316.2 & 17.801 & 311.7 & 313.2  & 17.787 & 2 & \\
0.09   & 356.5 & 312.0 & 17.841 & 380.9 & 311.0  & 17.825 & 2 & Mom$>$Coul \\
0.08   & 498.1 & 308.5 & 17.881 & 528.0 & 307.2  & 17.868 & 2 & Mom$\gg$Coul \\
\bottomhline
\end{tabular}
\label{tab:runs}
\end{table*}





\end{document}